Test Security in Remote Testing Age: Perspectives from Process Data Analytics and AI

Jiangang Hao[*] and Michael Fauss

Educational Testing Service

**Abstract**

The COVID-19 pandemic has accelerated the implementation and acceptance of remotely proctored high-stake assessments. While the flexible administration of the tests brings forth many values, it raises test security-related concerns. Meanwhile, artificial intelligence (AI) has witnessed tremendous advances in the last five years. Many AI tools (such as the very recent ChatGPT) can generate high-quality responses to test items. These new developments require test security research beyond the statistical analysis of scores and response time. Data analytics and AI methods based on clickstream process data can get us deeper insight into the test-taking process and hold great promise for securing remotely administered high-stakes tests. This chapter uses real-world examples to show that this is indeed the case.

---

**Introduction**

Test security is an integrated component of high-stakes assessments and is critical to ensure test results' validity, reliability, and fairness. Test security breaches can happen at different parts of a testing, such as testing sites, testing items, and testing processes. Therefore, a good test security solution often involves coordinated efforts around the prevention, detection, and remediation of security breaches. Among these efforts, data-based detection of security breaches plays an essential role in large-scale tests and the methods used in this area have evolved over the years as the available data changed.

Before the advent of digital technology, high-stakes tests were in paper-pencil format and administered in proctored sites. The data is limited to the final responses on the test papers. Statistical analyses were primarily used to detect test irregularities, such as similar response sequences due to answer copying. Since the late 1990s, the advance of digital technology has enabled computed-based and internet-based tests. For example, Educational Testing Service (ETS) started offering the computer-based edition of TOEFL in 1998 and the internet-based edition in 2005 (TOEFL Report, 2005). When tests are delivered through computers, collecting more data is possible. In addition to the final response data collected in paper-pencil tests, computer-based tests also capture response process data, such as response time, change of responses, navigation activities, and other clickstream activities. A number of statistical methods have been developed for detecting test irregularity, such as examining similar responses (Sotaridona & Meijer, 2003; Wollack, 1997; Cizek & Wollack, 2017; Kingston & Xlark, 2014; Wollack & Fremer, 2013; Haberman & Lee, 2017), item pre-knowledge (Lee and Lewis, 2021), score gains and anomalies (Skorupski & Egan, 2011; Wollack & Eckerly, 2017; Lee & von Davier, 2013), response time (van der Linden & Guo, 2008; Meijer & Sotaridona, 2006; Man,

Harring et al., 2018), and answer revision patterns (Primoli et al, 2011; van der Linden & Jeon, 2012; Wollack et al., 2015; Sinharary et al, 2017; Sinharay & Johnson, 2017 ). Data mining methods such as cluster analysis have also been suggested (Man et al., 2019). Readers should be aware that the above is an incomplete list of the relevant literature in the public domain. There are many methods and practices being used by different testing companies or organizations, but they are not in the public domain due to the sensitivity of test-security related research.

Since 2020, the COVID-19 pandemic has accelerated the implementation and acceptance of remotely proctored high-stakes tests. For example, ETS started offering TOEFL and GRE at-home editions in March 2020 (ETS News, 2020). In remotely proctored tests, test takers can take the tests from places (often at home) based on their convenience, as long as the site meets some requirements specified by the test administrator. While the flexible administration of the tests brings forth many values, it also raises concerns, such as security breaches and test interruption, which may impact the tests' validity, reliability, and fairness. Generally, we cannot control the hardware and environment in remotely proctored tests as tightly as in testing centers. This opens the doors for many possible unintended test-taking strategies to gain an advantage in the tests, creating many security concerns that do not exist for test-center administrations. However, addressing the hardware and environment issues often involves a nonnegligible burden for test takers. For example, bringing in a second or third camera can help to reduce the blind spots, but requires an extra cost for acquiring the cameras as well as improving the Internet infrastructure to meet the needs of the increased WIFI bandwidth to transmit the video data.

On the other hand, clickstream process data contain rich information about the test-taking process. Analyzing these process data allows us to develop indicators that can be used to flag suspicious test-taking behaviors to defend test security at a much lower cost compared to hardware enhancement. To extract useful information from the big clickstream data, one should go beyond the traditional psychometric/statistical methods and leverage techniques from data science, artificial intelligence (AI), and natural language processing (NLP), e.g., a computational psychometric way (von Davier, et al., 2021). ETS researchers conducted comprehensive research projects in this area (Hao, 2022), and some examples include clickstream data-based detection of remote computer access (Hao & Li, 2021), automated essay similarity detection (Novak, et al, 2022), using keystroke as biometrics to detect imposters (Choi et al., 2019), detection of AI generated essays (Yan et al., 2022), and the detection of retyping vs drafting using keystroke (Jiang et al.,, 2024; Zhang et al., 2022).

Despite the great promises, we want to caution that the current level of AI and analytics (even in the near foreseeable future) cannot completely replace human proctors for high-stake assessments, though AI and analytics can assist human proctors in improving their efficiency and accuracy in the proctoring process. Furthermore, joint consideration of innovations in item types, assessment design, and reporting is necessary to ensure the success of remote testing. In this chapter, we first outline the general methodology for using clickstream process data to support remotely proctored tests. Then we introduce how these new techniques can be applied in practice with some real-world examples.

**A General Roadmap**

Based on the data flow, there are four critical steps in using clickstream process data. Figure 1 shows these steps. The first step is to get all the relevant data and do a preliminary evaluation of their usefulness for different purposes. The second step is to transform the data and organize them in an appropriate format to feed into further analyses. The third step is to conduct extensive data mining work to develop features to characterize the process data. The last step is to connect the features developed in step 3 to scientific and business claims and present the results to stakeholders through interactive forms, such as dashboards. These four steps are not specific to the application for remotely proctored tests; they are the general steps of almost all data science projects. The steps are usually not a one-way process but an iterative one, meaning that the feedback from later steps often leads to adjustments (or reruns) of the earlier steps. As such, a close and agile collaboration among different parties involved is critical for moving the project forward.

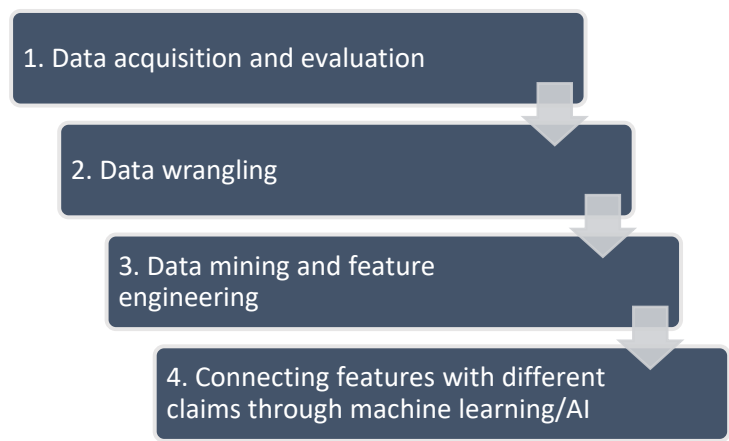

Figure 1. Critical steps in using process data to support remotely proctored tests.

For the data collection and evaluation step, it is important to differentiate two types of data, e.g., the outcome and process data. The outcome data refers to the final responses used for scoring, and there are well-established psychometric procedures to analyze them, which we will not cover in this chapter. The process data, however, capture the timestamped interactions between the test takers and the system. The process data could include the system telemetry log data and multimodal data such as video or audio. Traditional psychometric/statistical methods are not enough to fully extract the values of these process data, and new methods from data science, machine learning/AI, and NLP are generally needed. These process data and new methods lead to many security-related applications, such as detecting abnormal activities, excessive similarity of constructed responses, and AI-generated responses. Figure 2 is a schematic that outlines some of the major applications.

The methods for handling process data in test security applications could be roughly categorized into data analytics and AI (supervised machine learning). The data analytics approach refers to discovering meaningful features or patterns in data through data mining and making decisions based on these patterns, similar to unsupervised machine learning. As in unsupervised machine learning, there is generally no labeled data beforehand. Some typical applications of this approach in test security include the detection of similar essays or speech responses, the detection of suspicious test-taking behaviors, and zero-shot detection of AI-generated responses. The AI or supervised machine learning approach aims at building classifiers to map the feature representations of the data to a set of labels corresponding to different types of security violations. These classifiers will then be applied to new unlabeled data to detect potential security violations. Some applications of this approach in test security include, but are not limited to, imposter detection through biometrics, detection of copywriting, and detection of remote computer access. Even though we distinguish the two different approaches, it is worth noting that they are almost always used jointly in real-world applications. The following section introduces how these methods are applied in practice through some examples of empirical study.

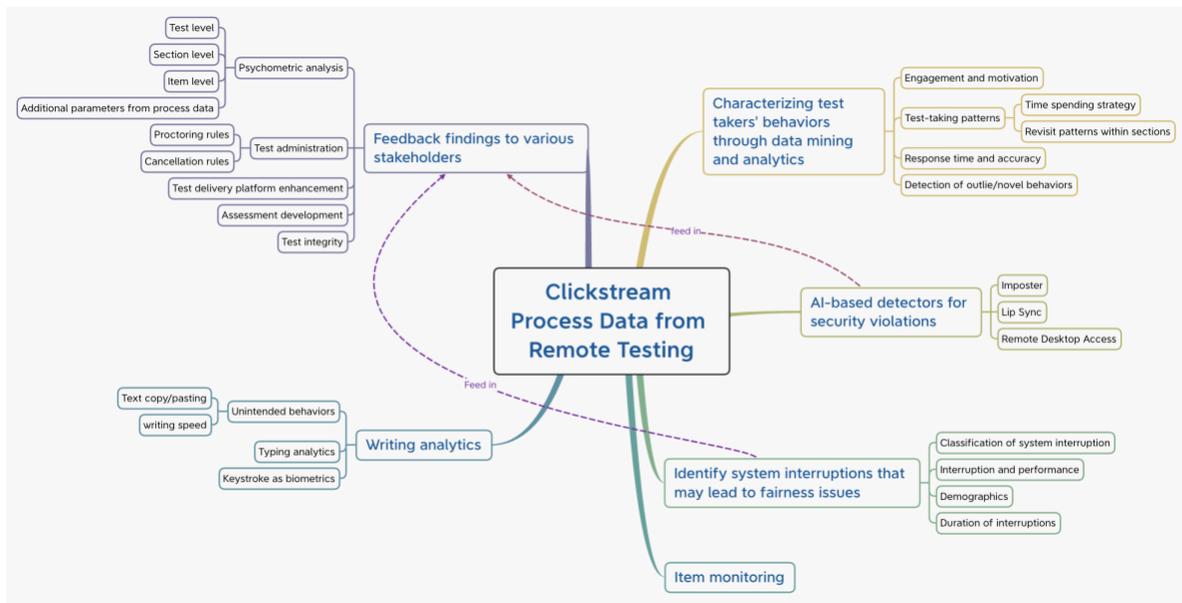

Figure 2. A summary of possible applications of process data in supporting remote testing

**Empirical Studies**

This section introduces two empirical studies highlighting how to use process data to support test security. Before we head to the details, we first introduce some basic terminologies to facilitate future discussions.

*Binary Classification and Evaluation*

One of the most important goals of test security-related research is to tell whether a given test session is cheating. As the outcome is binary (cheating or not), this line of research eventually leads to the development of various binary classifiers. In machine learning, the outcomes of binary classifiers are usually denoted as negative (no cheating) and positive (cheating). Based on this convention, a contingency table (also known as a confusion matrix) can represent the classification results, as in Figure 3.

|  | Predicted condition | |
|---|---|---|
| Total population = P + N | Positive (PP) | Negative (PN) |
| **Actual condition** Positive (P) | True positive (TP), hit | False negative (FN), type II error, miss, underestimation |
| Negative (N) | False positive (FP), type I error, false alarm, overestimation | True negative (TN), correct rejection |

Figure 3. Confusion matrix for binary classifier outcomes. Adapted from

https://en.wikipedia.org/wiki/Evaluation_of_binary_classifiers

From the contingency table (confusion matrix) of the outcomes, some important evaluation metrics can be established to quantify the performance of a binary classifier. For example, the False Positive Rate ($FPR = \frac{FP}{N}$) and False Negative Rate ($FNR = \frac{FN}{P}$) correspond to the type I and type II errors in statistical hypothesis testing, respectively. Here, the convention is that the negative case corresponds to the null hypothesis, and the positive case corresponds to the alternative hypothesis. In addition to $FPR$ and $FNR$, the True Positive Rate ($TPR = 1 - \frac{TP}{P} = 1 - FNR$) is another widely used evaluation metric.

The direct output of a binary classifier is often not a binary class but a continuous decision function. The algorithm usually decides a threshold on the decision function to assign positive or negative labels. As one changes the threshold, the positive and negative labels change accordingly, and so do the $FPR$ and $TPR$ of the classification result. The receiver operating curve (ROC) is introduced to characterize how $FPR$ and $TPR$ change as the decision threshold varies. Figure 4 illustrates the ROCs of good and poor classifiers. Two important statistics based on the ROC are widely used to quantify the performance of a classifier. One is the area under the ROC, usually denoted as AUC, which ranges from 0.5 for a random classifier to 1 for a perfect classifier. Generally, an AUC greater than 0.9 is considered an outstanding classifier, while an AUC from 0.7 to 0.8 is considered acceptable (Bradley, 1997).

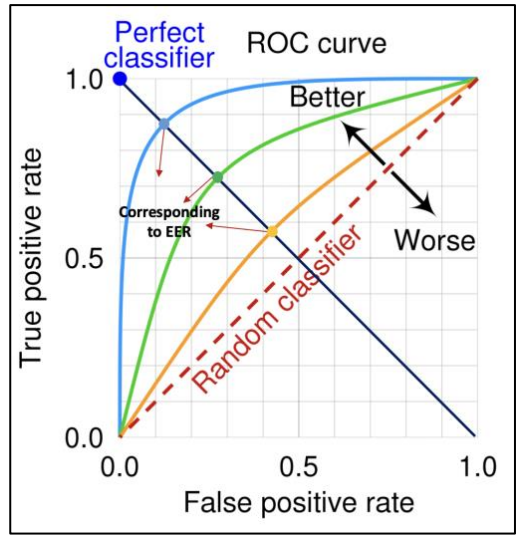

Figure 4. ROC curves of different classifiers adapted from

https://en.wikipedia.org/wiki/Receiver_operating_characteristic

Another important statistic from the ROC is the equal error rate, EER, which denotes the point where the *FPR* equals *FNR*. In Figure 4, three points corresponding to the EERs of the three classifiers are marked. Note that these points are not precisely EERs as the y-axis is not FNR but 1-FNR (e.g., TPR). EER is widely used in biometrics to characterize the classifier's performance; the smaller, the better. For example, the EER of the fingerprint is around 0.2%, and that of the iris is around 0.01%, though the actual numbers may be affected by the specific implementations (Walker, 2002).

In the following, we introduce two empirical studies to showcase how data analytics and AI/machine learning approaches can be used in the real world to support test security.

*Detection of ChatGPT-generated Essays*

The revolutionary advance of AI technology is changing learning and assessment in both positive and negative ways. There have been increasing concerns about test takers using AI-generated responses in writing tests, especially in remote testing. ETS researchers have conducted a systematic study to detect essays generated by a large language model, GPT-3 (Yan, et al., 2022). Two approaches were explored for the detection, one is a support vector machine classifier based on features generated by ETS' e-rater system (Attali & Burstein, 2006), and another is based on finetuning a pre-trained LLM, RoBERTa (Liu et al., 2019). The two methods achieved a classification accuracy of 95% and 99%, respectively (Yan, et al., 2023)

At the end of November 2022, OpenAI released another revolutionary AI product, ChatGPT (https://openai.com/blog/chatgpt/), which became an instant hit since its release. ChatGPT introduced a reinforcement learning with human feedback (RLHF) layer to handle user prompts/questions better than its predecessors, such as GPT-2 and 3, and can generate high-quality texts in a different context which has already put a serious challenge to the current way of evaluating students' writing. As a result, developing detectors that can detect ChatGPT-generated responses has become a hot research topic, and several detectors have been developed by different individuals and organizations in the public domain (Crothers et al., 2022; Slashdot, 2023; Tian, 2023). Most recently, OpenAI itself also released a detector for AI-generated texts (https://openai.com/blog/new-ai-classifier-for-indicating-ai-written-text/). In addition, there are many other detectors being developed but have yet been opened to the public, such as those developed at ETS and other organizations.

Three important upfront issues must be kept in mind while developing a serious detector of AI-generated essays (Hao, 2023). To clarify, we refer to a detector as "serious" if it is supposed to be used in a high-stakes application in which the detection outcome needs to be well justified and can lead to significant consequences. The first requirement is that a serious detector needs to have clear performance metrics, such as FPR and TNR, as discussed in the preceding subsection. A contrast sample must be specified to establish such metrics, e.g., against which sample the detector is trying to detect ChatGPT-generated essays. The second requirement is that the detector should be robust against reasonable human edits on top of the AI-generated texts, as this is how people use AI language models in the real world. The extent to which the detector should be robust against human edits is an open question and requires some consensus among the stakeholders in different use cases. Fundamentally, there is no way to tell for sure whether an AI or a human generated a text if enough paraphrases or revisions are applied. The third issue is the most challenging, namely, what evidence is necessary for one to confidently and justifiably claim that an essay is generated by AI in high-stakes applications. After all, it could cause more damages if someone is incorrectly accused of cheating using ChatGPT-generated responses. All these issues take work to address and generally require the consensus of stakeholders in specific use cases. In what follows, we introduce our work on developing a ChatGPT detector by keeping these three issues in mind.

  We chose a specific contrast sample when developing our detector. First, we randomly sampled two thousand essays from two prompts in the writing section of a high-stakes test. Then, we generated two hundred essays from each prompt using ChatGPT. These procedures guarantee a clear context for our detector. Furthermore, while we generated these essays from ChatGPT, we explicitly prompted the system to add some typos and grammar errors, by which we aim to mimic some level of human edits. Finally, once we had all the data ready, we conducted data analytics to identify features that could be used to detect the ChatGPT-generated essays.

We start with the simplest one of all possible features that can characterize the texts. As our goal is to detect whether a text is generated by AI, there is already a well-established quantity, perplexity (Mao, 2019), which characterizes how unlikely a sequence is generated by a given language model. The higher the perplexity, the less likely the sequence is from a given language model. When we have the essays from humans and ChatGPT, the simplest classifier is to compute the perplexity of the essays based on a given language model and then find out where to set the threshold as we know which essay is from ChatGPT. In the left panel of Figure 5, we show the density plot of the perplexity of essays from humans and ChatGPT, respectively. As one can read from the distribution, human-generated essays show a much broader range of perplexity, while ChatGPT-generated essays show a much narrower range. If we use the thresholds of the essay perplexity to control the two error probabilities of the decision function, the resulting ROC of the classification results is shown in the right panel of Figure 5. The results suggest that the essay perplexity alone already works very well for detecting ChatGPT-generated essays.

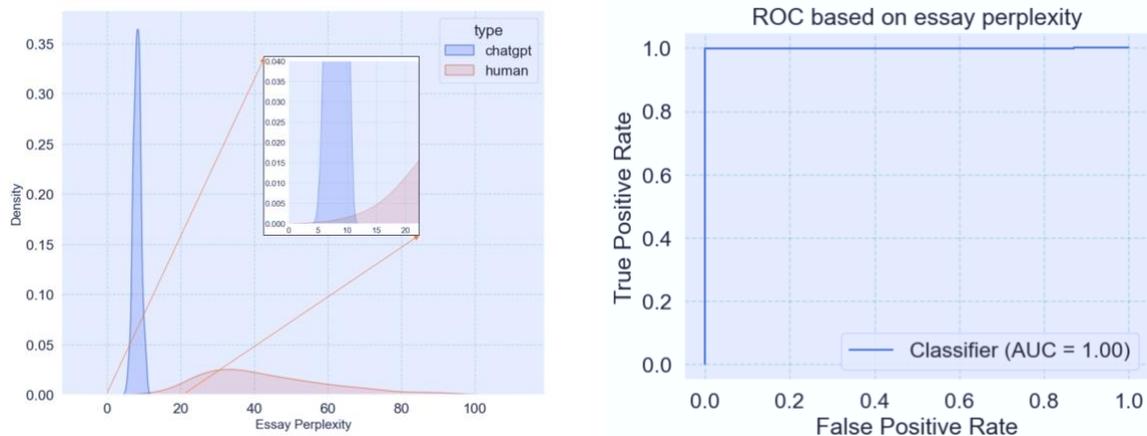

Figure 5. Left: distribution of the essay perplexity based on the GPT2 language model. Right: ROC of an essay perplexity threshold-based classifier.

It is worth noting that the above detector does not involve any supervised machine learning but simply applying a threshold on the perplexity of the essay, and it already performs very well, as measured by an almost perfect ROC. The essays used in the above analysis are relatively long, from 400 to 600 words on average. Can we still see the clear separation when the texts are shorter, e.g., at the sentence level? To verify this, we use the same essays but look at the distribution of the sentence perplexity. The results are shown in Figure 6. It is not unexpected that the detection power decreases at the sentence level, as the fewer words there are, the more difficult it is to tell whether a text was written by a human or an AI. For example, in the extreme case of one word, there is no way to tell if it is AI generated or not.

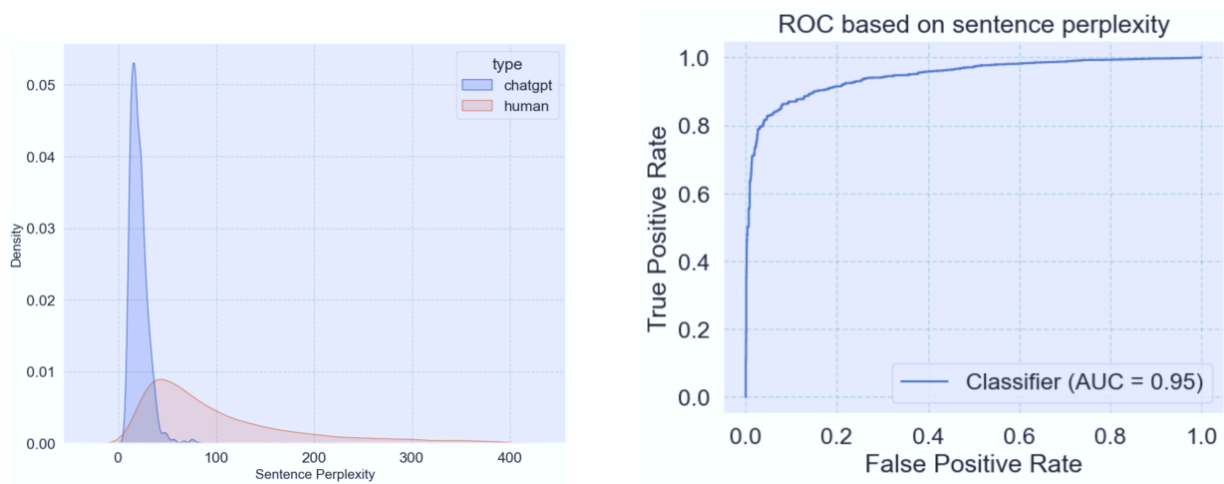

Figure 6. Left: distribution of the sentence perplexity based on the GPT2 language model. Right: ROC of a sentence perplexity threshold-based classifier.

Practically, it is often useful to have a guideline threshold on the length of the text, below which it becomes challenging to detect whether it is AI-generated or not. To get such a ballpark number, we first chose 200 human-written and 200 ChatGPT-generated essays. Then, we compute the perplexity of each essay by selecting the first N number of words, with N running from 10 to 400 at an increment of 10. We plot the mean and standard deviation of the perplexity in Figure 7. One can observe that when the text length is less than 100, it is generally difficult to separate AI-generated essays from human-written ones based on perplexity.

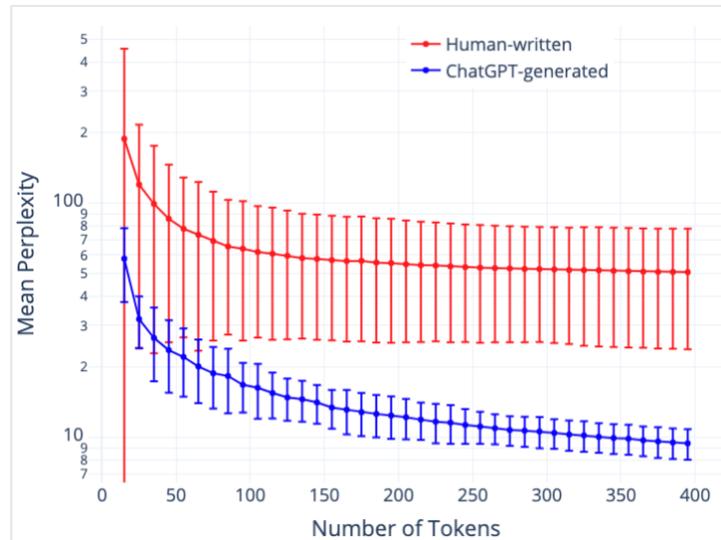

Figure 7. The mean and standard deviation of perplexity for texts of different length.

*Keystroke-based Detection of Suspicious Writing Behaviors*

In the preceding subsection, we introduced a simple detector of ChatGPT-generated essays. However, as we pointed out earlier, if a person makes many edits on top of AI-generated essays, we will surely lose the power of detection at a certain point, even with more robust detectors based on more language features and machine learning models. Therefore, additional information is needed to rescue the case, especially from another modality. Among them, the data? captured in the keystroke logs could provide valuable information about the writing process. It could be used to help with the detection when human edits are there. In the following, we show two applications of keystroke process data in test security, using it as a biometric to identify test takers and as a detector to uncover copywriting behaviors.

ETS researchers have extensively researched using keystroke data to understand students' writing process (Deane, 2014; Zhang & Deane, 2015). Recently, this line of keystroke capability has been applied to test security-related applications. For example, the possibility of using keystroke pattern as biometrics to identify test takers has been demonstrated (Choi, et al., 2021), and reliable detection of copywriting from draft writing based on keystroke patterns has been shown (Zhang, Hao, & Deane, in preparation). In the following, we will briefly introduce these two lines of research.

A person's writing process under similar writing environments shows unique patterns, which can be used as biometrics to identify the person. Some typical features to characterize the writing process include the number, latency, speed, and total time for specific typing events, initial and repeated backspace events, cut and paste events, and edits that involve jumps from one location to another. The features can also include measures about word edits, typo corrections, and the burst of text production. In addition, summary statistics of the time interval between adjacent letters (digraph) can also be used to characterize the writing process. Readers can refer to (Deane, 2014; Zhang & Deane, 2015) for more discussions of the development of features to characterize the writing process.

To demonstrate that keystroke features can be used to form biometrics to identify persons, we conducted an empirical study by choosing 3,110 repeated test takers (repeaters). We chose two essays from each repeater and created 3,110 repeater essay pairs. We also created 3,110 non-repeater essay pairs as a control by randomly pairing the essays from two non-repeaters. We extracted a set of writing process features for each essay based on the corresponding keystroke logs. Using the repeater essay pairs, we can also check how well the keystroke features are correlated in the two essays by the same test taker. Some highly correlated features are shown in Table 1.

| Feature name | Definition | Within-person correlation |
|---|---|---|
| inword_logIKI_median* | Median duration of in-word keystrokes, measured in log milliseconds | 0.95 |
| inword_logIKI_mean | Mean duration of in-word keystrokes in log milliseconds | 0.95 |
| wordinitial_logIKI_median* | Median duration of word-initial keystrokes in log milliseconds | 0.92 |
| append_interword_interval_logIKIs_mean | The mean log interkey intervals for keystrokes that add white space between words | 0.92 |
| wordinitial_logIKI_mean | Mean duration of word-initial keystrokes in log milliseconds | 0.92 |
| append_interword_interval_logIKIs_median* | The median log interkey interval for keystrokes that add white space between words | 0.91 |
| append_interword_interval_speed_median* | The speed of keystrokes that add white space between words, measured in characters per second | 0.91 |
| wordinitial_char_per_sec_median* | Median speed of typing the first character of a word, in characters per second | 0.91 |
| iki400_AppendBurst_len_mean | Mean length in characters of bursts of append keystrokes where no pause is greater than 400 milliseconds | 0.91 |
| iki400_AllActionBurst_len_mean | Mean length in characters of bursts where all keystrokes count as part of the burst, and bursts end on pauses longer than 400 milliseconds | 0.90 |
| initial_backspace_char_per_sec_median* | The median speed of the first in a series of backspace actions, measured in characters per second | 0.90 |
| iki200_AppendBurst_len_mean | Mean length in characters of bursts of append keystrokes where no pause is longer than 200 milliseconds | 0.90 |
| initial_backspace_logIKI_median* | The median log interkey interval for backspace actions that appear first in a series of backspace actions | 0.89 |

Table 1. Highly correlated keystroke features from the same test taker (adapted from Choi et al., 2021).

Based on these keystroke features, one could develop a classifier to detect whether an essay pair is from the same or a different person. To do this, for each essay pair, we created a distance feature vector by calculating the Euclidean distance between the same keystroke features from each essay in the essay pair. This way, the problem becomes a typical supervised machine-learning task. After comparing different algorithms, we found that the Gradient Boosting Machine (Friedman, 2001) works best for our data. The resulting classifier achieved performance with an equal error rate of 4.7%. This finding provides empirical evidence that keystroke patterns in writing tests could be used as a complementary biometric measure, adding an additional security layer. For more details on this study, we refer readers to our research paper (Choi et al., 2021).

Another important application of keystroke analytics in test security is to detect copywriting from draft writing. The cognitive process of draft writing involves four subprocesses, proposer, translator, transcriber, and evaluator (Hayes, 2012). Figure 8 shows a diagram of the four cognitive subprocesses and their meanings. Features from the keystroke process data can be mapped to these cognitive subprocesses (Deane et al, 2018; Zhang & Deane, 2015). For example, the initial pause before writing corresponds to the proposer subprocess.

On the other hand, in copywriting, where a test taker copies some existing texts and types them to respond to the writing items, the cognitive process is different. For example, the proposer and translator subprocesses could be completely obsolete, and the evaluator subprocess will be different accordingly. The different cognitive processes could leave their traces in the keystroke process data logs.

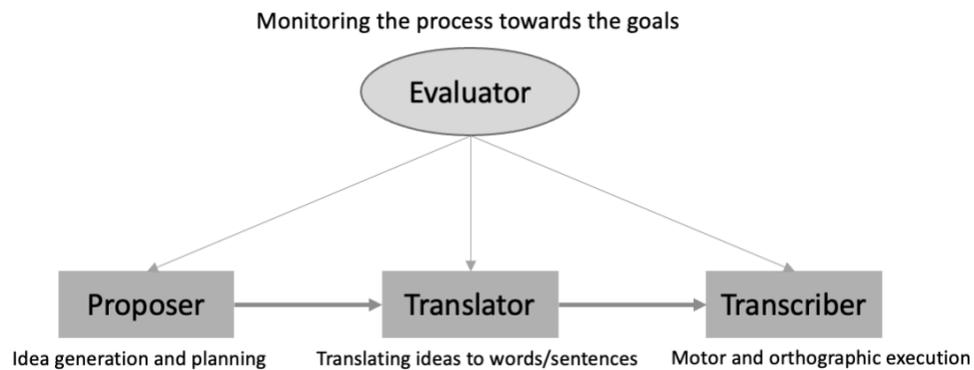

Figure 8. The cognitive process of writing based on Hayes, 2012.

Zhang, Deane, and Hao (2022) reported that a machine learning classifier based on keystroke features can detect copywriting well from authentic writing in a controlled experiment, with an accuracy of over 95%. Jiang et al. (2024) showed that the accuracy of detecting non-authentic texts using keystroke features from large-scale assessment could reach over 90%. If test takers copy from ChatGPT-generated essays into writing assessment, the keystroke features could provide additional information to complement the text perplexity to further enhance the detection accuracy.

The above two keystroke-related applications make it clear that writing process data, when properly used, could have important implications for test security. Keystroke data capture the timestamped typing activities, thus providing important information about the process of how texts were developed. This information is not attenable through traditional NLP approaches, and therefore provides a unique handle for detecting AI-generated responses with human revisions.

**Summary**

Remote testing has become a crucial means of administering high-stakes tests during the COVID-19 pandemic, with its growing popularity and adoption worldwide. Remote testing offers several benefits, such as convenience, accessibility, and flexibility for test-takers. It also presents cost and time savings and easy scalability for test administrators, as the logistic challenges associated with testing centers are significantly reduced. As a result, remote testing will likely continue to be an important way of administering high-stakes tests even after the pandemic.

However, remote testing also raises many security concerns, as it is more challenging to maintain security measures of a test when it is taken from a not fully uncontrolled site. In this chapter, with three empirical examples, we have shown that data analytics and AI methods applied to clickstream process data allow us to gain deeper insight into the test-taking process and lead to new methods for detecting anomalies and suspicious behaviors that may indicate cheating. This line of research is indispensable for securing remote testing and should represent an important future direction of test security research.

**Acknowledgment**

This work was funded by ETS Research Allocation and Test Security Initiative.**Reference**

Attali, Y., & Burstein, J. (2006). Automated essay scoring with e-rater® V. 2. The Journal of Technology, Learning and Assessment, 4(3).

Bradley, A. P. (1997). The use of the area under the ROC curve in the evaluation of machine learning algorithms. Pattern recognition, 30(7), 1145-1159.

Breiman, L. (2001). Random forests. Machine Learning, 45(1), 5-32

Choi, I., Hao, J., Deane, P., & Zhang, M. (2021). Benchmark Keystroke Biometrics Accuracy from High-Stakes Writing Tasks. ETS Research Report Series, 2021(1), 1–13.

Cizek, G. J., & Wollack, J. A. (Eds.). (2017). Handbook of quantitative methods for detecting cheating on tests. New York, NY: Routledge.

Crothers, E. et al. (2022). "Machine Generated Text: A Comprehensive Survey of Threat Models and Detection Methods." ArXiv abs/2210.07321

Deane, P. (2014). Using writing process and product features to assess writing quality and explore how those features relate to other literacy tasks. ETS Research Report Series, 2014(1), 1–23.

Deane, P, Steck, F., Roth, A., Lewis, M., Litz, A., Richter, T., & Goswami, V. (2018). Behavioral Differences Between Retyping, Drafting, and Editing: A Writing Process Analysis. ETS Research Memorandum, RM-18-06

ETS News, (2020, March). https://news.ets.org/press-releases/ets-introduces-at-home-solution-for-toefl-ibt-test-and-gre-general-test-amid-coronavirus-pandemic/

Friedman, J. H. (2001). Greedy function approximation: a gradient boosting machine. Annals of Statistics, 1189–1232.19